# Triggering and guiding electric discharge by a train of UV picosecond pulses combined with a long UV pulse

A.A.Ionin, S.I.Kudryashov, A.O.Levchenko, L.V.Seleznev, A.V.Shutov, D.V.Sinitsyn,
I.V.Smetanin, N.N.Ustinovsky, V.D.Zvorykin

P.N.Lebedev Physical Institute of Russian Academy of Sciences
53 Leninsky pr., 119991 Moscow, Russia
Phone/Fax: 7(499)7833690, e-mail: aion@sci.lebedev.ru

**Abstract**

Non-self-sustained electric discharge and electric breakdown were triggered and guided by a train of picosecond UV pulses overlapped with a long free-running UV pulse of a hybrid Ti:Sapphire-KrF laser facility. Photocurrent sustained by this train is two orders of magnitude higher, and electric breakdown distance is twice longer than those for the discharge triggered by the long UV pulse only.

Plasma channels produced by laser radiation in atmospheric air or some other gases are of great interest for many fundamental problems and technical applications. There are triggering and diverting of lightning [1,2], directing of microwave radiation to overcome its original divergence [3,4], laser-driven acceleration and guiding of electrons [5] among them. In contrast to early experiments with $CO_2$ laser pulses of microsecond length [6], where opacity of dense plasma produced via avalanche ionization limited the length and continuity of the channel, approaches based on the use of UV [1,2,7,8] or femtosecond [9,10] laser pulses can produce long-distance partially ionized tracks in air (or other gas) due to multiphoton ionization either with or without filamentation of radiation. Since primary photoelectrons are quickly recombined with positive ions and attached to the molecular oxygen (during the period of time ~10-50 ns), additional influence of longer laser pulse is anticipated to keep the electron density for much longer time [1]. There are several papers [11-13] in which combination of single fs and single ns pulses resulted in plasma revival and improved triggering and guiding electric discharges. However, they have dealt not with UV pulses but with near IR fs and visible or near IR- ns pulses. Application of a long train of ultrashort UV laser pulses or combination of such a train with a long UV pulse seems to be the most attractive for creation and further supporting of plasma channels [14]. In this paper we demonstrate that non-self-sustained electric discharge and electric breakdown are triggered and guided by a train of picosecond UV pulses overlapped with a long free-running

UV pulse of a hybrid Ti:Sapphire-KrF laser facility. Photocurrent sustained by this train is two orders of magnitude higher, and electric breakdown distance is twice longer than those for the discharge triggered by the long UV pulse only.

The hybrid Ti: Sapphire – KrF laser system [15] that emitted a single subpicosecond TW pulse [16] was modified in such a way to produce combined ultrashort and long UV laser pulses. The layout of experiments is shown in **Fig. 1**. UV ultrashort pulse of 0.5-mJ energy, 100 fs duration at wavelength λ=248.4 nm was emitted by the frequency-tripled Ti:Sapphire front-end facility Start 248 manufactured by Avesta Project Ltd. This single pulse or a train of ultrashort pulses with period of 5ns produced by the single pulse going through the ring cavity at the exit of front-end facility were first amplified in double-pass e-beam-pumped KrF preamplifier BERDYSH [15]. After passing 6 m length spatial filter the laser beam was collimated by the lens (*F*=3.3 m) into a parallel beam and was injected into main KrF laser amplifier GARPUN [15] that was equipped by a confocal unstable cavity with magnification factor M=6. This cavity with double-pass time of 17 ns was formed by a rear concave high reflectivity mirror (*R*=6 m) and partially transparent convex output mirror. The foci of both the concave and convex mirror were matched.

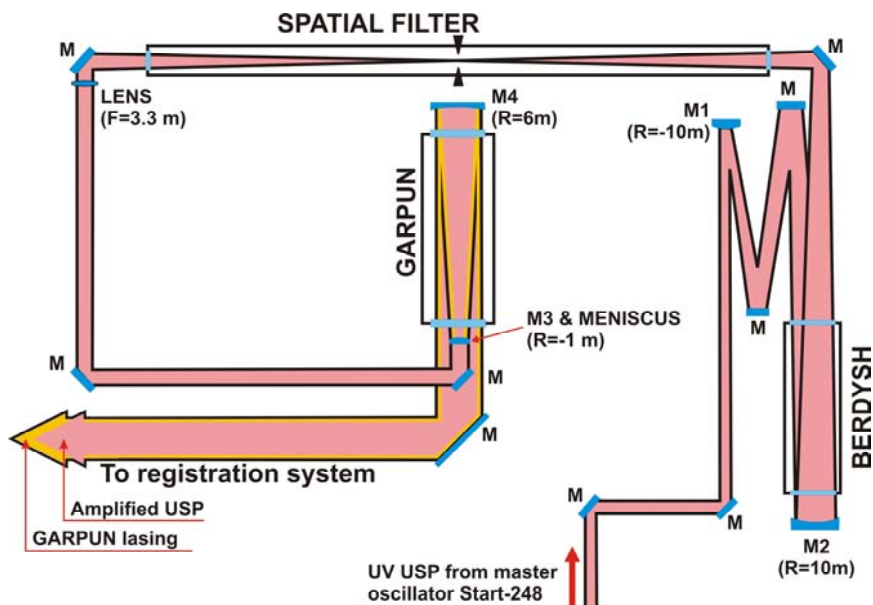

**Fig.1** Optical scheme of producing a train of UV ultrashort pulses (USP) combined with a long UV pulse.

Because of multi-pass amplification of the injected ultrashort pulse (or injected train of ultrashort pulses with period of 5 ns) in this cavity and free-running lasing of the KrF laser with this cavity, the laser system emitted a train of picosecond (because of group velocity dispersion in optical windows) UV pulses with period of 17 ns (in the case of single pulse injection) or 5ns (in the case of ultrashort pulse train injection) superimposed over a long pulse of tens of ns duration emitted by aforementioned KrF laser (see **Fig.2a,** where the total laser pulse obtained under injection of the train of ultrashort pulses, is presented). The ultrashort pulse intensity in diagram of **Fig.2a** is ~$10^3$ times less than that of the real ultrashort pulse because of a long response time ~1ns of a vacuum photodiode integrating ultrashort pulse signal. The total energy of the combined laser pulse depended on reflectivity of the output mirror and was varied between 10 and 30J. The maximum peak power of UV ultrashort pulse was estimated to be 0.2 TW.

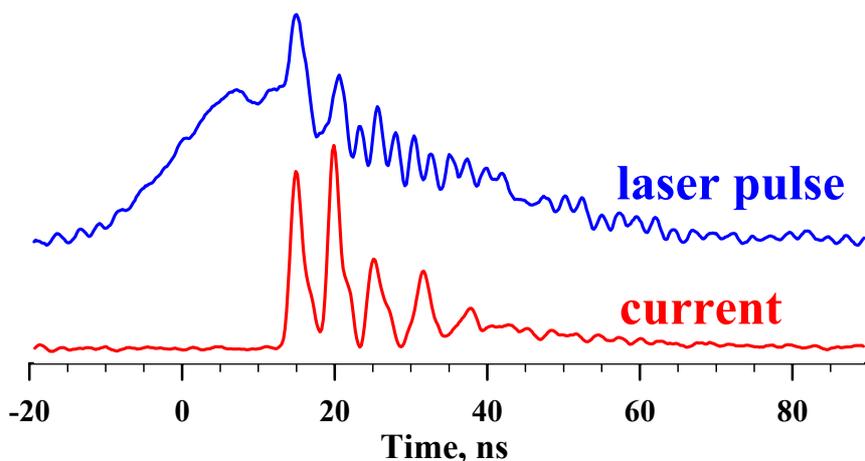

**Fig.2** Combined laser pulse obtained under injection of a train of ultrashort pulses (a) and photocurrent sustained by this pulse (b).

Three different experimental schemes were used for study of triggering and guiding electric discharge by a train of UV picosecond pulses combined with a long UV pulse. In the first one this UV radiation was focused by a concave mirror with focus length ~8.0 m in between ring electrodes situated 20 cm from each other through the high-voltage ring electrode along the inter-electrode gap and produced a plasma channel. High-voltage 5 – 22 kV was applied between the electrodes. UV ultrashort pulse triggered and sustained a non-self-sustained electric discharge, the voltage not being enough for electrical breakdown (a self-sustained discharge) of the inter-electrode gap. Photocurrent signal corresponding to injection of the ultrashort pulse train measured on matched impedance input $R_{osc}$= 50 Ω of an oscilloscope is presented in **Fig.2b**. This photocurrent signal consists of a sequence of short current pulses of ~2 ns length following with period of ~5 ns. (These short current pulses followed with period of 17 ns when a single ultrashort pulse was injected). Comparing **Fig.2a** and **Fig.2b** one can see that photocurrent amplitude is close to zero for the beginning of free-running UV pulse and sharply increases for ultrashort pulse. When focusing only smooth free-running laser pulse with the same energy (under blocked ultrashort pulse injection), the photocurrent amplitude was 100 times less than that for ultrashort pulse. Such a difference in photocurrent amplitudes is connected with nonlinear nature of air ionization depended on laser intensity. The photocurrent signal was approximately the same as the geometric focal point of the focusing mirror was moved along the optical axis and inter-electrode gap within distance of ~1.0 m, which enabled us to estimate the plasma channel length as ~1.0 m.

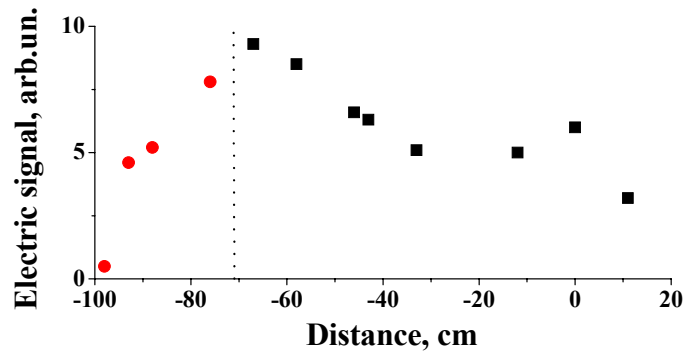

**Fig.3** The dependence of electrical signal (arb.un) on the electrodes position (a distance from the geometric focus point). The vertical dashed line indicates the electrodes position where laser beam is limited by the diaphragm. Circles are obtained with this limitation. Laser beam goes from the left to the right.

It should be noted that the peak ultrashort pulse power exceeded the critical power of self-focusing for $\lambda$=248-nm radiation (~0.1 GW [17]) by three orders of magnitude. Multiple filamentation of ultrashort pulse took place, which was observed as "hot points" even in the near-field pattern. The length of plasma channel formed under the filamentation of ultrashort pulse under its focusing was measured by using the second scheme for a single subpicosecond ultrashort pulse obtained under direct amplification of 100 fs laser pulse of the front-end Ti: Sapphire facility in the two e-beam pumped KrF laser amplifiers. (Final amplifier GARPUN was not equipped with the unstable cavity). Contrary to the first scheme the single laser pulse was focused by a mirror system with a numerical aperture of $7 \cdot 10^{-3}$ in between a pair of spherical electrodes of 20 mm in diameter perpendicular to the axis of inter-electrode gap. To prevent the electrodes from illumination by laser radiation and photo-effect emergence, a diaphragm of 10-mm in diameter was placed in front of the 11-mm inter-electrode gap. As plasma was created in the inter-electrode gap by the focused UV radiation, the capacitor formed by the electrodes was recharged, and an electrical signal was observed with an oscilloscope. The electrode system was moved along the optical axis, and the maximum amplitude of the electrical signal proportional to the electron amount in the plasma was recorded. The dependence of the electrical signal on the electrodes position is shown in **Fig.3**. The UV laser pulse with energy of 0.25 J and peak power of 0.5 TW, a few orders of magnitude higher than critical power ~0.1 GW for self-focusing, propagates from the left to the right undergone by multiple filamentation and forming a bunch of plasma channels. Zero of the X axis corresponds to the position of the geometric focal point. The vertical dashed line indicates the electrodes position where the laser beam is limited by the diaphragm. Circles are obtained with this limitation. The diagram of **Fig.3** demonstrates that the total length of the plasma channel formed by multiple filaments of ultrashort UV pulse is about one meter. Plasma cross-section and electron density distribution along the plasma channel were apparently not homogeneous. The electric signal amplitude is lower close to the geometric focal point. No photo-ionization signal at all was detected at the distance just ~10 cm behind the geometrical focal point. For distances longer than 70 cm in front of geometric focus the amplitude of electrical signal decreased due to the fact that some laser radiation was limited by the diaphragm.

The formation of long conductive plasma channels sustained by combined UV laser pulses during tens of ns gives us an opportunity to efficiently trigger and guide an electrical breakdown of long air gap. First experiments on triggering and guiding electric breakdown of not very long inter-electrode gap were carried out by using the third scheme. Pulsed UV radiation was focused in between electrodes of 10 mm in diameter along the inter-electrode gap. These experiments demonstrated that free-running UV pulse triggered electric breakdown of 4.0 cm length inter-electrode gap with applied voltage of 50 kV, with self-sustained discharge starting ~5 μs after the laser pulse and its trajectory not depending on laser beam. At the same high-voltage and laser output energy a train of UV picosecond pulses combined with a long UV pulse triggered and guided electric breakdown along the laser beam through the distance of 7.0 cm with breakdown time-delay relatively the laser pulse at least two orders of magnitude shorter than in former case.

Therefore, a combined UV pulse was demonstrated to produce plasma channels of much higher conductivity, trigger longer electric discharge guiding it along the laser beam as compared to free-running long pulse of the same energy and duration.

The research was supported by Grant # 097007 of US Air Force European Office of Aerospace Research and Development through partner Project 4073 of the International Scientific Technological Center and partially by Russian Foundation for Basic Research through Projects # 10-02-01477, 11-02-01100, 11-02-01414, 11-02-01524 and 11-02-12061.